# ESCAPE

Preparing Forecasting Systems for the
Next generation of Supercomputers

## D3.1
## Recommendations and specifications for data scope analysis tools

Dissemination Level: public

This project has received funding from the European Union's Horizon 2020 research and innovation programme under grant agreement No 67162

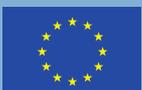
Funded by the
European Union

Co-ordinated by

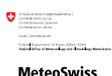 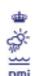 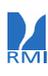 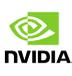 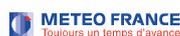 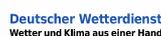 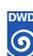 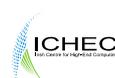 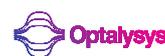 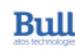 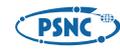 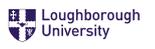

# ESCAPE

**E**nergy-efficient **S**calable **A**lgorithms
for Weather **P**rediction at **E**xascale

Author **Cyril Mazauric, Erwan Raffin, David Guibert (Bull)**

Date **19/12/2017**

# Table of Contents

## Figures

## Tables

# 1 Executive Summary

In today's computer architectures, moving data is considerably more time- and energy consuming than computing on this data. One of the key performance optimizations for any application is therefore to minimize data motion and maximize data reuse. Especially on modern supercomputers with very complex and deep memory hierarchies, it is mandatory to take data locality into account.  Especially when targeting accelerators with directive systems like OpenACC or OpenMP, identifying data scope, access type and data reuse are critical to minimize the data transfers from and to the accelerator. Unfortunately, manually identifying data locality information in complex code bases can be a time consuming task and tool support is therefore desirable.

In this report we summarize the results of a survey of currently available tools that support software developers and performance engineers with data locality information in complex code bases like numerical weather prediction (NWP) or climate simulation applications. Based on the survey results we then recommend a tool and specify some extensions for a tool to solve the problems encountered in an NWP application.

The tools analysed for this survey are:

- DDT debugger by Allinea/ARM
- Extrae and Dymemas by the Barcelone Supercomputing Centre (BSC)
- Perf-mem
- Totalview and MemoryScape debuggers by RogueWave
- OpenSpeedShop
- MemAxes by Lawrence Livermore National Laboratory (LLNL)
- Redux Valgrind plugin

While many of the above tools provide some functionality useful for the aforementioned use cases, none of the tools readily offered the capability to track data scope and accesses throughout an application. As it turned out, none of the tools were capable of combining static and dynamic analysis while keeping the link with the original source code.

Based on the survey results, our recommendation is to use the Extrae and Dymemas tools developed at BSC. Their functionality is closest to what is needed for data scope analysis in NWP applications. While some critical features are still missing to turn this into a truly useful tool for the end user, we are in communication with BSC and expect the remaining features to become available in a future version of the toolchain.

# 2 Introduction

## 2.1 Background

ESCAPE stands for Energy-efficient SCalable Algorithms for weather Prediction at Exascale. The project develops world-class, extreme-scale computing capabilities for European operational numerical weather prediction and future climate models. ESCAPE addresses the [ETP4HPC](ETP4HPC) Strategic Research Agenda 'Energy and resiliency' priority topic, promoting a holistic understanding of energy-efficiency for extreme-scale applications using heterogeneous architectures, accelerators and special compute units by:

- Defining and encapsulating the fundamental algorithmic building blocks underlying weather and climate computing;
- Combining cutting-edge research on algorithm development for use in extreme-scale, high-performance computing applications, minimizing time- and cost-to-solution;
- Synthesizing the complementary skills of leading weather forecasting consortia, university research, high-performance computing centers, and innovative hardware companies.

ESCAPE is funded by the European Commission's Horizon 2020 funding framework under the Future and Emerging Technologies - High-Performance Computing call for research and innovation actions issued in 2014.

The work presented in this report directly contributes to achieving top-level objectives 2 and 3.

### 2.2 Scope of this deliverable

#### 2.2.1 Objectives of this deliverable

In this deliverable we provide a survey of the currently available state-of-the-art data scope analysis tools in the context of NWP applications. In addition, we offer a recommendation and specifications for a memory analysis tool easing both porting and optimization of ESCAPE dwarfs on high-end processors and accelerators such as GPUs.

#### 2.2.2 Work performed in this deliverable

The goal of this first task is to survey the available data scope tools and gain some first-hand experience on these tools. The main selection criteria for these tools was the capability to properly determine read and write accesses of variables in complex programs in addition with the corresponding source code, a key information to determine the optimal locality for a variable in systems with highly non-uniform memory access times.

The list of readily available tools includes: DDT (Allinea Debugger), Extrae and Dymemas (BSC tools), Perf mem, TotalView Memoryscape (Roguewave Debugger), OpenSpeedShop, Redux Valgrind plugin, MemAxes/Mitos.

Most of these tools have been tested and used on basic applications. The only exceptions are the Redux Valgrind plugin which is in a very prototypical state since 2005 and not maintained anymore and MemAxes/Mitos which was unusable on systems available at Bull.

Among the tested tools, only the tools developed at BSC seem to be close to enabling memory access and pattern analysis, including the relation with the source code and other performance metrics. This task is therefore ongoing as BSC tools are still under development (not publicly available and tested in collaboration with the POP CoE). Thanks to our test cases, some features and improvements have been added to the next release of the memory tool.

Finally, there are ongoing discussions with the developers of the Portland Group compiler (PGI) about using the compiler infrastructure for data dependency analysis to obtain data locality information at compile time. While much of the desired information is available internally to the compiler, a reporting tool for this information would still need to be developed and is not yet part of a product roadmap.

### 2.2.3 Deviations and counter measures

None.

## 3 Problem specification

### 3.1 Porting and optimizing for accelerators: the GPU case

One of the challenges when moving an application to accelerators via OpenACC is to determine the data access patterns, the scope of variables and the reuse patterns. Especially with current generation GPUs and their connectivity to the host memory system, excessive data transfers between host and GPU can significantly hamper the performance. Unfortunately, large code bases as encountered in the Climate/Weather community exhibit often complex access patterns and it is difficult to know a-priori which data needs to be staged to the device at which point. Future generation GPUs will bypass the low bandwidth PCI express bus and data access to main memory from the GPU can occur at speeds exceeding the CPU access speed to host memory. However, even in these future situations, it will be beneficial to minimize the page migration between host and GPU memory to minimize the memory access latency.

We are therefore looking for tools that will provide the following information:

- At what point is the data first accessed?
- When and in what routines is the data accessed again?

Granularity of this information should at least be at the array level, ideally at the level of individual elements. The tools should be capable of identifying accesses to the same data even if it is obfuscated by variable renaming, passing to subroutines or by extraction of subarrays. Both static and run-time analyses are of interest.

### 3.2 General problem statement

The primary question addressed by a data scope analysis tool is how and where in the code a data structure is allocated and accessed. This information, essentially the dataflow graph, can help to identify relevant memory access optimizations, especially in large code bases such as NWP applications. Moreover, extracting the memory patterns and feedback from the memory sub-system using dynamic analysis on a set of relevant input data helps to understand the application behaviour on a particular hardware.

An additional feature of interest in the optimization process is to detect data races and unsafe reads from shared variables. This can help to ensure that a loop nest is parallelizable and help to determining the privacy level of a shared variable. This feature has been implemented for example in GeCoS as "ompVerify" and integrated into the Eclipse IDE and its CDT (C/C++ Development Tooling) to verify C/C++ programs with OpenMP pragmas, aiming to provide real-time static verification for OpenMP programmers [1]. Unfortunately, FORTRAN – which is commonly used in the NWP community – is not supported, thus it has been excluded from the survey.

## 4 Tools presentation

In this section we will present the tools surveyed and tested for this report. For each tool we will mainly focus on the functionality related to the memory monitoring.

### 4.1 DDT: Allinea debugger

DDT is a parallel and multithread debugger which proposes an intuitive GUI. DDT is mainly used to debug your application but can also be used to monitor memory.

#### 4.1.1 Memory profiling functionalities

DDT's memory monitoring is enabled by setting a breakpoint at the source line of interest. The application execution will stop at this line and you will be able to check all needed information: memory usage balanced (see the left part of Figure 1), number of allocation calls, deallocation calls (see the right part of Figure 1), value of each variables, etc.

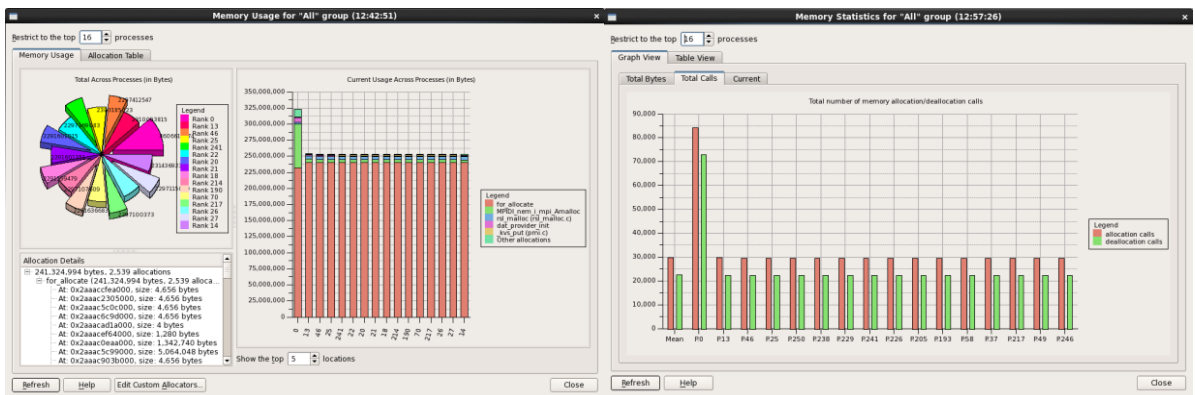

*Figure 1 : DDT - Overall memory usage*

At each time, the value of all variables could be visualized easily (Figure 2)

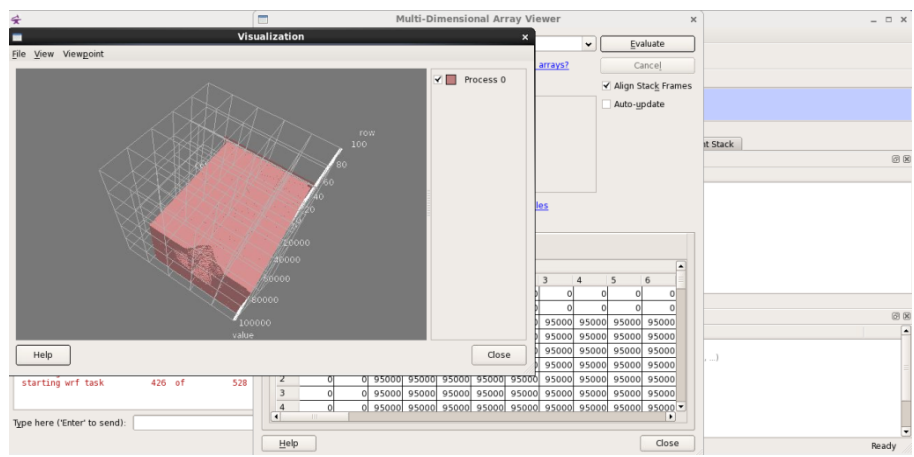

*Figure 2 : DDT - Variable visualization*

An interesting feature is that DDT enables the user to see where a table has been allocated irrespective of the breakpoint position. (right button then Pointer details - Figure 3)

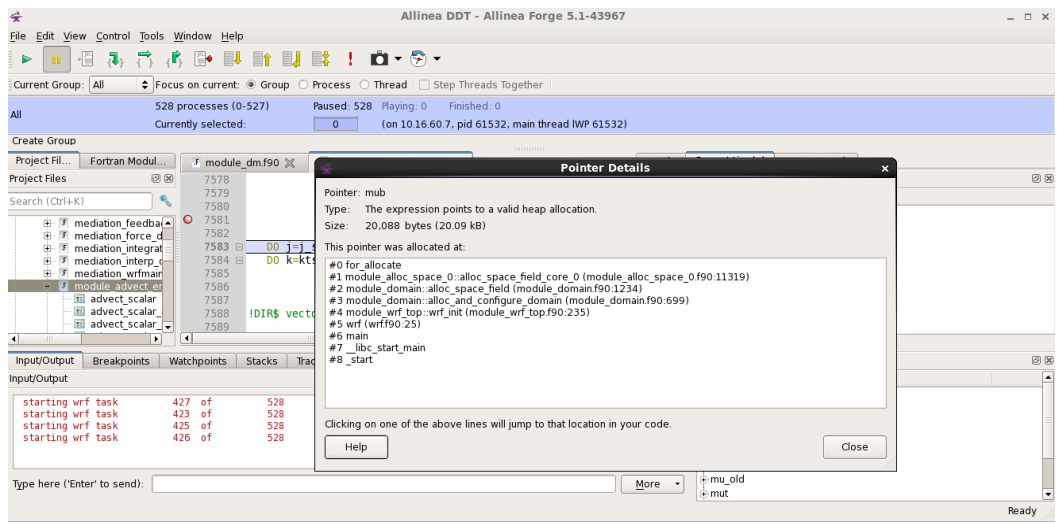

*Figure 3 : DDT - Pointer Details*

### 4.1.2 Pro & cons

DDT is easy to use and has a very intuitive GUI. We can easily follow, step by step, the value of each variable and detect where these variables have been allocated. The monitoring is done during the execution which allows one to choose precisely all parameters of the run.

Your application should be compiled with the following options –g –O0. Compiled with –O0 can slightly modify the application behaviour but is necessary to give all information to DDT.

One of the interesting features of DDT is that we can automatically detect where in the code a variable has been allocated. Unfortunately, this feature only works for dynamically allocated variables. For others variable, DDT will show you where this variable has been defined.

## 4.2 Perf mem

Perf is a lightweight performance monitoring tool that is included in the Linux kernel. It can instrument CPU counters and perform static and dynamic tracing.

### 4.2.1 Memory profiling features

For memory profiling purpose, perf tool provides the "*perf mem*" command. This command profiles memory access frequency (load and/or store operations, the "and" is not available in the first versions).

Basic usage:

ESCAPE 2016

```
$ perf mem record –o perf.data ./myApp
$ perf mem report –i perf.data
```

A high level memory report is given by:

```
$ perf mem report –i perf.data –s mem
```

```
Samples: 1M of event 'cpu/mem-loads/pp', Event count (approx.): 43753995
Overhead        Samples  Memory access
  33.16%        1396159  L1 hit
  29.78%          96017  L3 hit
  18.74%          19000  Local RAM hit
  18.13%          60612  LFB hit
   0.19%           2605  L2 hit
   0.01%             11  L3 miss
   0.00%            168  Uncached hit
   0.00%              1  Remote Cache (1 hop) hit
```
*Table 1 - High level memory report*

The command below produces an interactive report as shown in Table 2.

```
$ perf report –i perf.data
```

```
Samples: 1M of event 'cpu/mem-loads/pp', Event count (approx.): 2857401496
Overhead  Command  Shared Object      Symbol
  29.21%  hydro    hydro              [.] trace
  27.63%  hydro    hydro              [.] riemann
  12.48%  hydro    hydro              [.] updateConservativeVars
   6.37%  hydro    hydro              [.] slope
   5.15%  hydro    hydro              [.] qleftright
   4.55%  hydro    hydro              [.] gatherConservativeVars
   4.52%  hydro    hydro              [.] cmpflx
   4.44%  hydro    hydro              [.] constoprim
   2.80%  hydro    hydro              [.] equation_of_state
   2.19%  hydro    hydro              [.] compute_deltat
   0.27%  hydro    hydro              [.] ToBase64
   0.01%  hydro    [kernel.kallsyms]  [k] update_wall_time
```
*Table 2 - perf report "load" example*

Annotated source code can be displayed as shown in Table 3 for the first line ("trace" function), where figures in first column report the percentage of samples for function "trace" captured for that instruction.

```
         |                   cc = c[s][i];
         |                   csq = Square(cc);
```

```
           |            r = q[ID][s][i];
           |            u = q[IU][s][i];
           |            v = q[IV][s][i];
   3.64    |        vmovsd (%r15,%r12,8),%xmm3
           |        vmovsd %xmm3,0x1c8(%rsp)
           |            p = q[IP][s][i];
   3.91    |        vmovsd (%r11,%r12,8),%xmm7
           |           dr = dq[ID][s][i];
   3.91    |        vmovsd (%r10,%r12,8),%xmm3
```

*Table 3 - perf report annotate trace example*

Another useful feature of perf that can be used to profile memory operations such as allocations/deallocations is the ability to define custom probes. Unfortunately, this feature requires root permission and is therefore of limited use to the end user.

#### 4.2.1.1 Availability on Red Hat Enterprise Linux

According to Red Hat [2], this feature is available as of Red Hat Enterprise Linux 6.6. Red Hat Enterprise Linux (RHEL) 6.6 minimum kernel version is 2.6.32-504. It can be noticed that, as the kernel (or part of it) can be updated, the "*perf mem*" command can be available on older RHEL version than 6.6.

Moreover, depending on the processor generation and type, Perf may or may not acess performance monitor counters (PMCs).

### 4.2.2 Pro & cons

Perf mem is provided as part of the Linux kernel and allows profiling a broad range of hardware and software events. It provides memory profiling information with different granularity, from synthetic view to annotated assembly code. In addition, it supports optimized and OpenMP applications.

But, perf mem does not answer the core problem of data locality, at least not at the user level. An additional challenge is that it's only supported on recent kernel versions, and is typically not available in currently operational systems. However, this problem will be mitigated with newer systems being installed.

## 4.3 Extrae and Dymemas : BSC tools

The BSC tool set provides low-overhead detection of memory access patterns and their time evolution [3]. More precisely, BSC has extended the folding mechanism firstly described in [4] to the memory reference samples, and to collate all the metrics (source code, memory references and node-level performance) in one report per profiled region.

To capture referenced addresses, BSC enhanced framework uses a combination of two monitoring tools to generate a single trace-file that includes hardware counter performance metrics, call-stack references and data references. These tools are Extrae (that uses PAPI) to collect hardware performance metrics and perf to collect memory references from either load or store instructions. The graphics below show the classic usage of these tools.

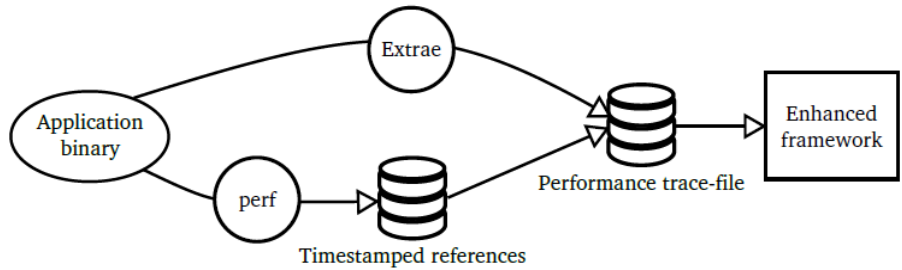

*Figure 4 - Combination of Extra and perf to generate performance trace-file including memory references*

### 4.3.1 Memory profiling functionalities

Thanks to these enhancements, BSC added a new feature which relies on the ability to report time-based memory access patterns in addition to source code profiles and performance bottlenecks.

In the article [3] BSC demonstrates how these tools deal with the memory access profile.

This can be demonstrated using the classic STREAM memory bandwidth benchmark. In order to use the tool, small source code modifications are necessary: a static C array definition needs to be replaced with a malloc dynamic allocation and the loop body needs to be instrumented.

The graphic below shows how EXTRAE is able to monitor the execution of the STREAM and to summarize the memory access with a clear picture.

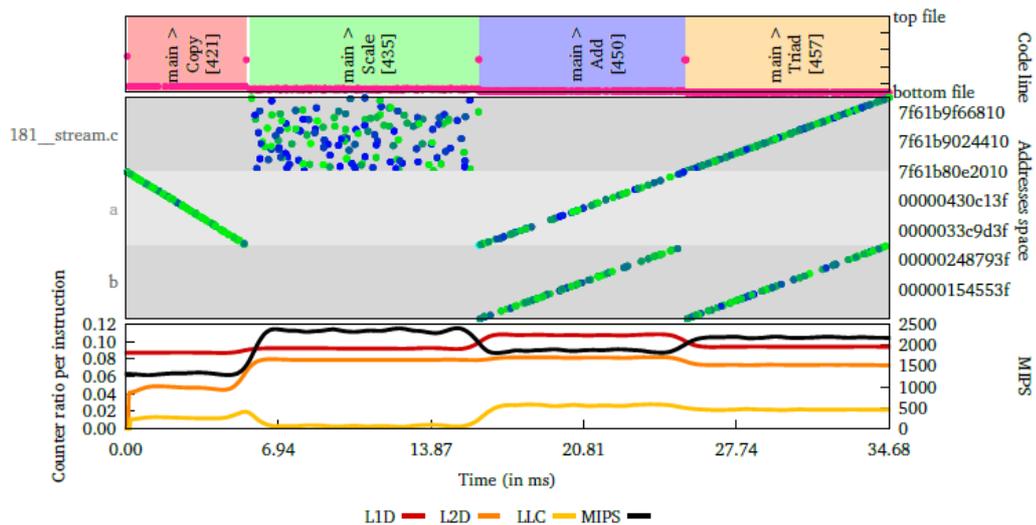

*Figure 5 - Analysis of the modified Stream benchmark. Triple correlation time-lines for the main iteration: source code, addresses referenced and performance.*

The first part of this graphics (top), shows the active routine within the source code. The second part (middle) shows the address space for each variable which has been allocated and accessed. The third part shows the achieved instruction rate within the instrumented region, as well as the L1D, L2D and LLC cache miss ratio per instruction.

In the same article, another profile has been done with CGPOP and BIGDFT to show how EXTRAE behave with a real application. More recently, LULESH memory access has been profiled [5].

#### 4.3.2 Pro & cons

With BSC tools it is easy to extract profiling from real application. The usage of the monitoring tools, DYMEMAS, has been improved and several classic profiles are proposed by default.

The source codes need to be modified to instrument the region of interest. In the same way as DDT, EXTRAE can automatically detect all variable which have been allocated dynamically and proposed to follow among the time execution the memory access patterns.

### 4.4 Totalview and MemoryScape: RogueWave debugger

Totalview is a graphical debugger for serial and parallel code. Totalview is mainly used to debug your application. With Totalview, RogueWave privdes a memory debugger: MemoryScape.

#### 4.4.1 Memory profiling functionalities

MemoryScape allows developers to watch for memory leaks and monitor memory usage and heap allocations while an application is running.  Enhanced facilities enable developers to monitor heap memory, view memory usage, locate memory leaks, track memory events, and show corrupted memory.

To monitor your application with TotalView, you can add a breakpoint at the line where you want to work or let the application go to the end (Figure 6).

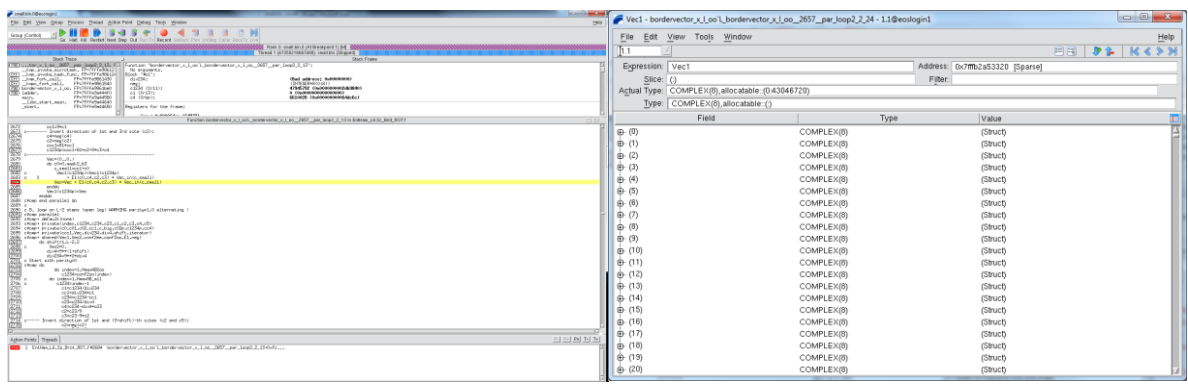

*Figure 6 : Totalview - Overall memory usage*

Once the application stops, you can double click on the targeted table to see its characteristics (Type, values …). From this point it is not possible to see where this table has been allocated.

Memoryscape could be opened from Totalview to begin the memory debugging.

Heap status Graphical report section allows the user to visualize the memory mapping and to see which variables have been allocated. You can sort variables by size in order to work on the most memory consuming (Figure 7). Double clicking on the desired table will open the code where the variable has been allocated.

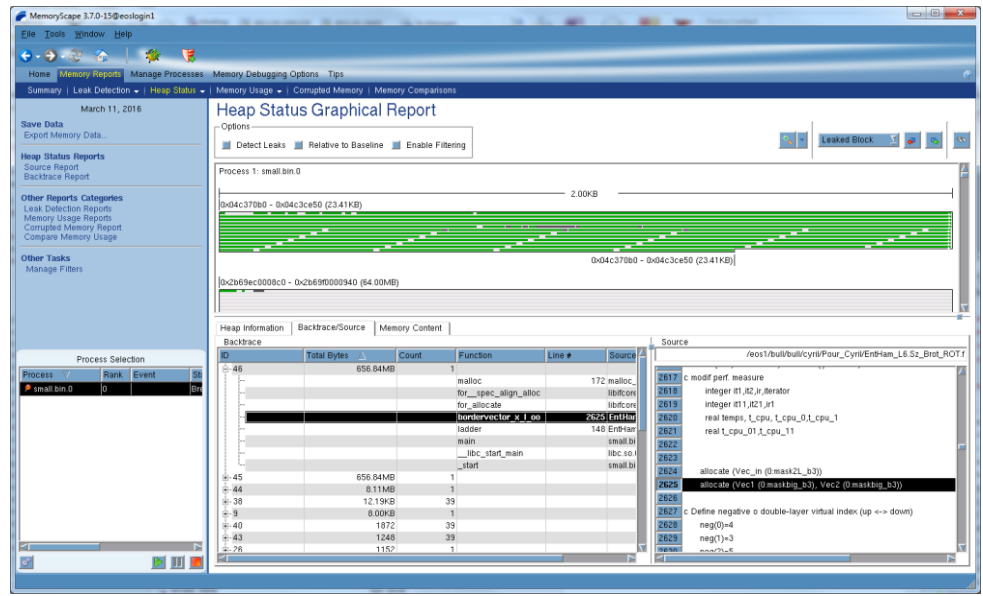

*Figure 7 : Totalview - Variable visualization*

MemoryScape could be used directly without Totalview.

### 4.4.2 Pro & cons

Totalview is easy to use; we can easily follow, step by step, the value of each variable. After recording information, RogueWave proposes to use ReplayEngine to let you move forward and backwards within these previously executed instructions.

Finally, MemoryScape could be used to detect memory issues and will show you the real memory mapping and the code where each table has been allocated.

Your application should be compiled with the following options –g –O1. Compiled with –O1 will, slightly, modify the application behavior but is necessary to give all information to Totalview and MemoryScape.

Thanks to MemoryScape you can determine the list of tables which have been allocated by the complete application and see the code where it happened, but this feature works only for variables which have been allocated dynamically.

Unfortunately, it is not possible, from the line of the code where a variable is used, to know where this variable has been allocated.

### 4.5 OpenSpeedShop

OpenSpeedShop is an open source multi-platform Linux performance tool which is targeted to support performance analysis of applications running on both single node and large scale systems.

#### 4.5.1 Memory Analysis Techniques

The OpenSpeedShop (version with CBTF collection mechanisms) supports tracing memory allocation and deallocation function calls in user applications. It is able to provide synthetic information (as shown in Table 4) and also a detailed call tree of each call to allocation and deallocation functions (as shown in Table 5).

```
openss>>expview

 Exclusive      % of   Number     Min       Min      Max      Max Total Bytes  Function (defining location)
  Mem Call     Total       of  Request Requested  Request Requested   Requested
  Time(ms)      Time    Calls    Count     Bytes    Count     Bytes
178.929739 89.959577   144650                                                   free (libmpi.so.12.0: i_rtc_hook.c,45)
 19.942766 10.026521    35390       24         1       20 426712512 10776777728 __libc_malloc (libc-2.17.so)
  0.027651  0.013902       48       24       150       24       816       23184 realloc (libmpi.so.12.0: i_rtc_hook.c,70)
```

*Table 4 - Synthetic memory allocation and deallocation*

```
openss>>expView -v CallTrees,FullStack mem1 -m max_bytes,retval,size1,size2

      Max           Function  Size Arg  Size  Call Stack Function (defining location)
Requested          Dependent            Arg
    Bytes       Return Value
                                              _start (hydro)
                                              > @ 562 in __libc_start_main (libmonitor.so.0.0.0: main.c,541)
                                              >>__libc_start_main (libc-2.17.so)
                                              >>> @ 517 in monitor_main (libmonitor.so.0.0.0: main.c,492)
                                              >>>> @ 219 in main (hydro: main.c,140)
                                              >>>>> @ 218 in allocate_work_space (hydro: hydro_funcs.c,166)
426712512    0x7efd2a8e3010  426712512     0  >>>>>>__libc_malloc (libc-2.17.so)
```

*Table 5 - Call tree of the biggest memory allocation*

#### 4.5.2 Pros & cons

The memory analysis feature of OpenSpeedShop provides synthetic and detailed memory allocation and deallocation information. The finest granularity is the function/procedure. It is possible to get the returned value (i.e. pointer to the base address) of each allocation and the function where this allocation is done.

As this tool only focuses on memory allocations and deallocations, it doesn't give any information on data access, thus it can't answer the problem.

### 4.6 Portland Group Compiler (PGI compiler)

The PGI compiler (www.pgroup.com) has a long tradition of providing quality Fortran support for HPC applications. It is also one of the leading OpenACC compilers, pushing the forefront of compilation for heterogeneous systems.

As pointed out in Section 2, a tool providing the desired capability can work both using static dependency analysis as well as runtime data analysis.

The PGI compiler contains powerful static dependency analysis engines that work within a single compilation unit. For example, when an OpenACC user braces a piece of code with a $ACC KERNELS region, the dependency analysis will determine which variable will need to be migrated to the GPU to make them available on the device. This is exactly the capability that's needed. However, one of the shortfalls is that this mechanism only works within a single compilation unit, and cannot take into account dependencies in functions called within this routine.

We are currently investigating the suitability of the current Unified Memory support offered by the PGI compiler for the purpose of data scope analysis.

### 4.7 Intel Advisor

Intel Advisor is a SIMD vectorization optimization and shared memory threading assistance for C, C++, C# and Fortran software. It supports SSE, AVX, AVX2 and AVX-512 instructions for Intel processors (Phi included), generated by Intel, GNU and Microsoft compilers. OpenMP and MPI applications are supported as well. It is available on Linux and Windows operating systems as a standalone GUI tool or plugin for Microsoft Visual Studio.

#### 4.7.1 Performance advisor

Intel Advisor can provide data-driven vectorization optimization recommendations. The analysis of an application considers mainly investigation of loops to answer questions such as: Whether they should be threaded or vectorised first? Is the vectorisation efficient enough? Are there any dependencies within the loop that prevent vectorisation? What are the trip counts and memory access patterns? Moreover, the user is provided with detailed information regarding instruction set (traits and data types) that are invoked within the analysed loop. The report provides useful information about each loop with respect to vectorisation recommendations and potential vector issues, including inefficient memory access patterns, external dependency, RAW (read after write) or WAR (write after read) dependencies. This tool also suggests loops which should process larger amounts of data in order to be vectorised. An example of such dependencies presented by the survey report is presented in Figure 8.

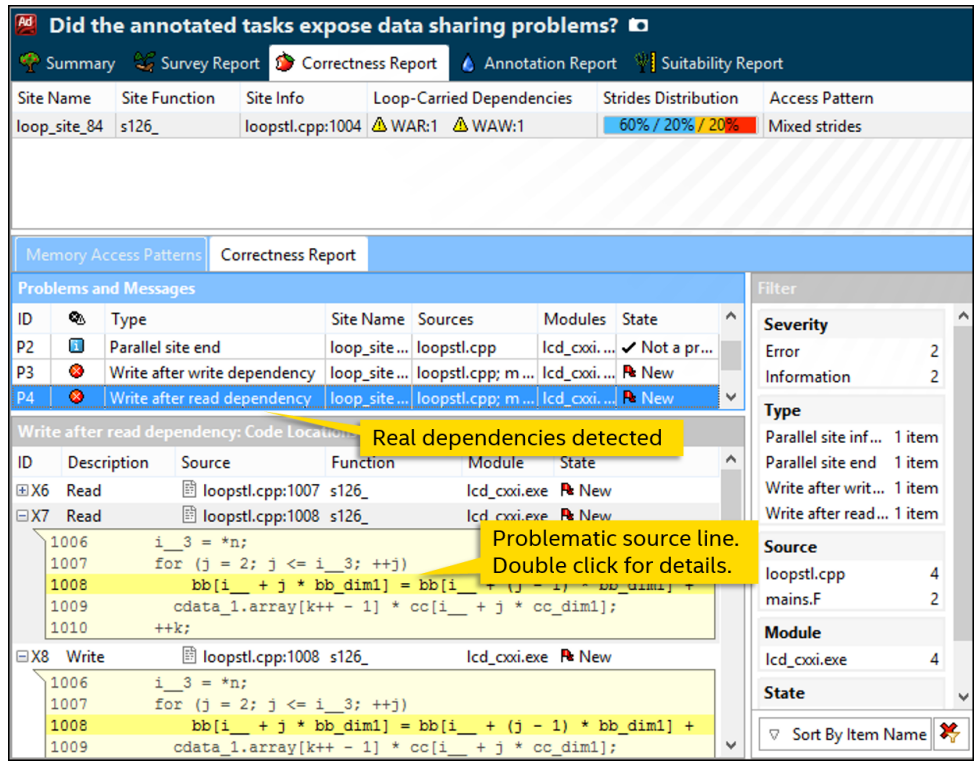

*Figure 8: Intel Advisor data dependencies report example.*
*Source: software.intel.com*

For an application to be analysed no specific code instrumentation is required. The user needs to run survey analysis via command line advixe-cl command, e.g.

```
 advixe-cl -collect survey -project-dir
/root/intel/advixe/projects/TCO639_ITER100_F100 -- mpirun -n 1 hwloc-bind --cpubind
socket:0.core:0-13 -- /root/Project/ESCAPE/software/dwarf-D-spectralTransform-
sphericalHarmonics/builds/dwarf-D-spectralTransform-sphericalHarmonics/bin/./dwarf-
D-spectralTransform-sphericalHarmonics-prototype1 --config TCO639.json
```

and then additional survey to collect e.g. trip counts and FLOPs:

```
advixe-cl     -collect     tripcounts     -flops-and-masks     -project-dir
/root/intel/advixe/projects/TCO639_ITER100_F100 -- mpirun -n 1 hwloc-bind --cpubind
socket:0.core:0-13   --   /root/Project/ESCAPE/software/dwarf-D-spectralTransform-
sphericalHarmonics/builds/dwarf-D-spectralTransform-sphericalHarmonics/bin/./dwarf-
D-spectralTransform-sphericalHarmonics-prototype1 --config TCO639.json
```

The last step is to prepare final report for further analysis in GUI, e.g.:

```
advixe-cl -report survey -show-all-columns -csv-delimiter=semicolon -format=csv -
project-dir /root/intel/advixe/projects/TCO639_ITER100_F100
```

Figure 9 illustrates an example screenshot from Intel Advisor GUI when performing Spherical Harmonics dwarf analysis.

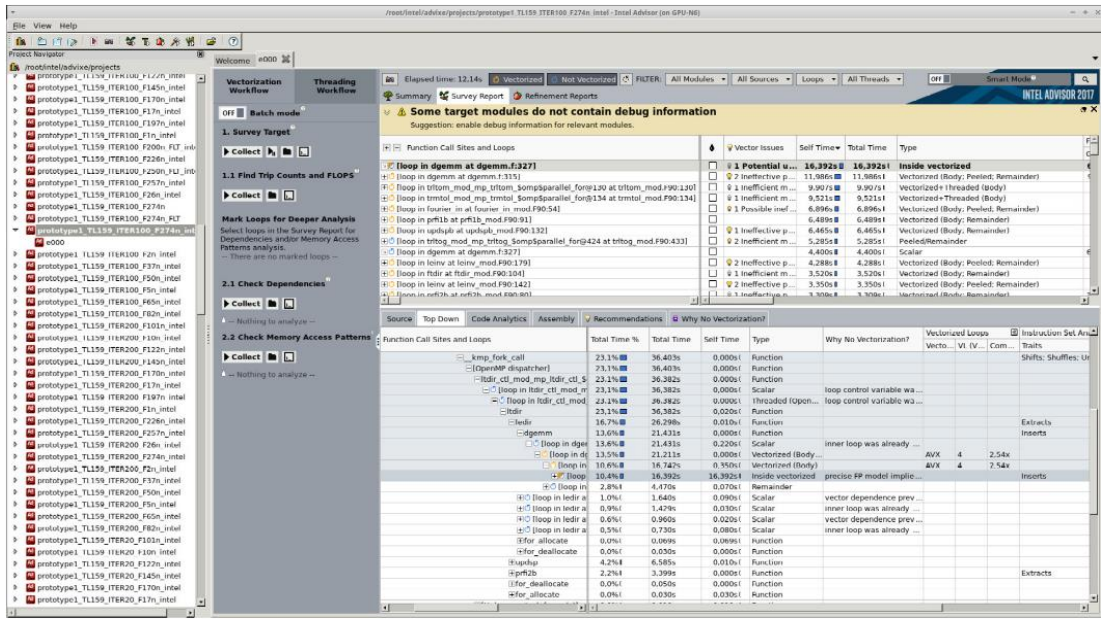

*Figure 9: Intel Advisor screen during SH dwarf analysis*

### 4.7.2 Memory access patterns

A useful feature of Intel Advisor is a memory access pattern survey. This tool identifies loops with contiguous, non-contiguous and irregular access patterns. It provides aggregated statistics on how frequently each pattern took place in a given source loop, maps each stride to specific objects in the code, and suggests recommendations how to improve access patterns. Additionally, the distance in bytes metric gives maximum distance between min and max memory address values. Figure 10 presents an example analysis of the memory access pattern – the 2/3 of analysed loop contains contiguous access pattern, while 1/3 has irregular access pattern.

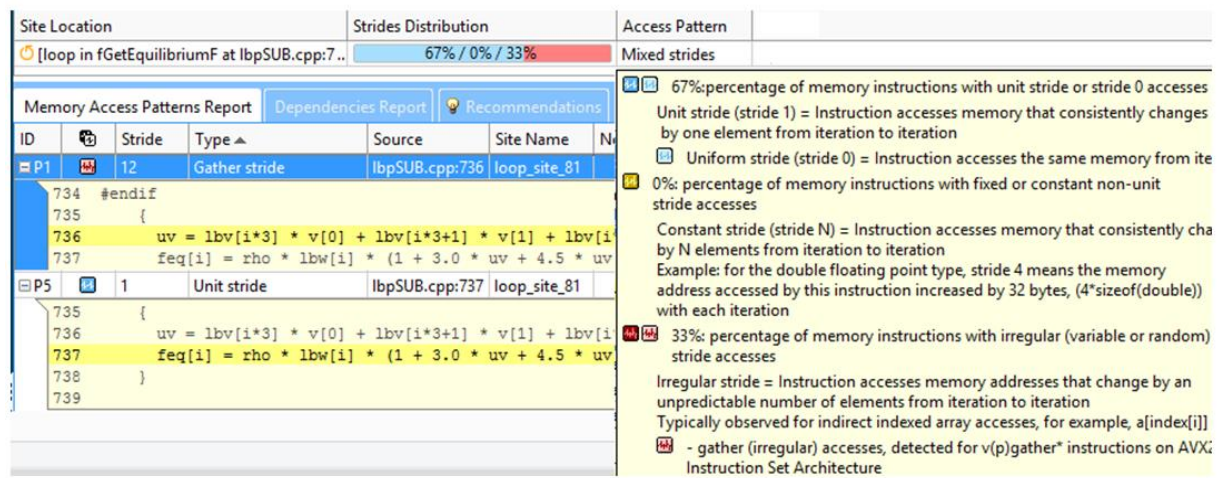

*Figure 10: Memory access patterns analysis with Intel Advisor.*
*Source: software.intel.com*

#### 4.7.3 Pros & cons

The Intel Advisor tool is a valuable profiler to enhance efficiency of the software. It provides data scope analysis in terms of improving data access patterns, loop vectorisation and loop/variable dependencies. Although it is not strictly a memory profiler tool, it allows one to perform improvements w.r.t. data locality and spatiality. A wide range of supported compilers (Intel, GNU, Microsoft), supported languages (C, C++, C#, Fortran), and parallel programming paradigms (OpenMP, MPI) make this tool even more attractive. The only downside is that it comes with just a 30-day free trial, however many HPC centres offers this tool to their users on free of charge basis.

This tool is used to provide performance models for NWP dwarfs, as described in D3.2 Performance models.

## 5 Survey conclusion

In summary, a collection of tools with memory analysis features have been surveyed. None of the tools available satisfies our requirements, namely an analysis on a per variable basis on data scope and access (read/write) for a given source code.

The BSC tools seem to currently offer the best fit for the desired capability in terms of dynamic analysis. Two restrictions should be noticed: data allocated on the stack is not trackable (as other similar tools) and the link with the source is still in progress. On the desired capability which consists of helping the developers to understand the scope of a data object, a strong compilation infrastructure is mandatory.

The PGI approach based on their static dependency analysis engines is promising in that way. More thorough investigations of these tools are ongoing as described in the next section and new tools are being surveyed as they emerge. The data scope analysis directly supports the performance models in task 3.2.

Finally, the ideal tool is based on a combination of static and dynamic analysis to deal with memory optimization and porting to accelerators.

## 6  Application of BSC tools on an ESCAPE dwarf

Collaboration with BSC via the POP Centre of Exellence [6] has been started concerning the memory analysis of three dwarfs, especially the *dwarf-d-spectraltransform-sphericalharmonics* and is still in progress. Indeed, a problem of unresolved symbols doesn't allow making the link between memory addresses and source code. Nevertheless, a first draft report has been received on the base of three runs (using 24, 48 and 96 cores)

This report contains the following information:

- Memory used for three different address regions (high, middle and low addresses).
- Time spent, and number of accesses for the different regions.
- Behaviour of the memory accesses over time (Figure 11).
- Average latency in cycles to get memory from the different memory levels (L1, Line Fill Buffer - LFB, L2, L3, DRAM).
- Time spent, and number of accesses for the different memory levels.
- Caller levels in the callstack for memory accesses.
- Memory hierarchy location to identify memory regions accessed by the application (see Figure 12).

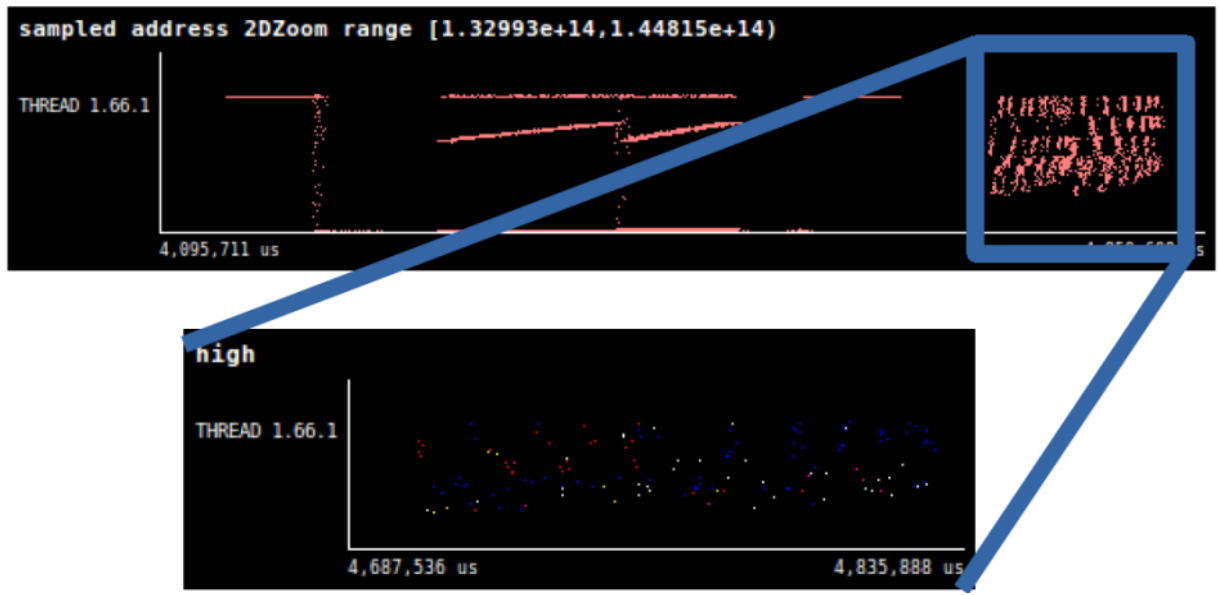

*Figure 11 - Structure for the high addresses for the 96 rank run with a zoom on a more random access pattern than other regions*

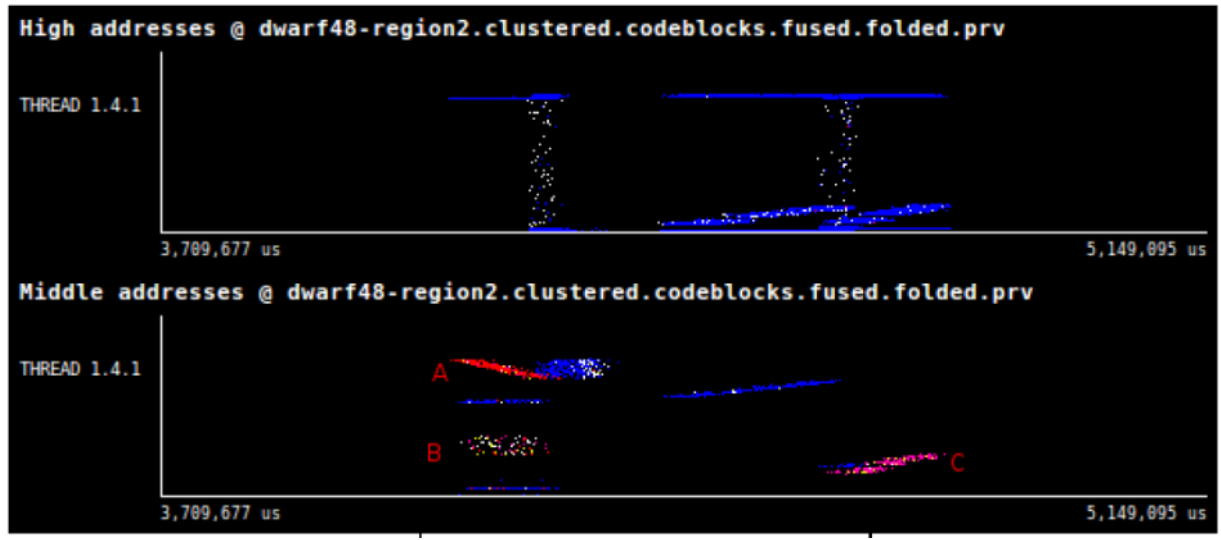

*Figure 12 - Memory hierarchy location of sampled loads. Top image shows the high memory addresses, and middle image show the middle memory addresses. Different regions have been identified and marked A-C.*

This first draft is promising as it gives some interesting observations and hints on the memory profile of the dwarf even if the link with the source code is missing.

## 7 Recommendations and specifications for memory scope analysis tool

The needs are twofold when optimizing and porting large applications:

- First, developers need to pay attention to how, where and when data are allocated, structured and used; this means to extract data scope and more generally to perform dataflow analysis. The control flow graph (CFG) is given by a static analysis of the code thanks to a compiler infrastructure and can be refined by dynamic analysis to add runtime and data dependent information. The latter extracts the dataflow graph for a particular input data set.
- Second, when running an application with a representative input data set on a specific hardware system, developers need to understand how the memory subsystem behaves to be able to determine how it can be used more efficiently. This feature can be provided by analysing memory accesses at runtime using a profiler and analyser tool.

The combination of static and dynamic analyses, in addition with the ability to make the link with the source code, are the key features required to help developers at porting and optimizing large application (e.g. NWP) in order to improve data locality thus time-to-solution and energy-to-solution.

First of all, providing statically at source code level the results of dataflow analyses is the first specification that helps when dealing with transfer between a host and an accelerator. This feature required a strong compiler infrastructure together with a source-to-source approach.

Then, providing at runtime a profile of the memory subsystem and its associated analysis at source code level greatly helps in finding bottlenecks in the application in terms of data locality and thus providing optimization hints (loop transformation, etc.).

## 8   Conclusion

This deliverable has presented the outcomes of the survey on available data scope analysis tools for the analysis of data usage and locality in large numerical weather prediction (NWP) applications. Thanks to this survey, the resulting recommendations and specifications for such a tool have been sketched.

## 9 References

[1] V. Basupalli, T. Yuki, S. Rajopadhye, A. Morvan, S. Derrien, P. Quinton and D. Wonnacott, "ompVerify: polyhedral analysis for the OpenMP programmer," *OpenMP in the Petascale Era,* pp. 37-53, 2011.

[2] "RED HAT ENTERPRISE LINUX - PERFORMANCE TUNING GUIDE," 2017. [Online]. Available: https://access.redhat.com/documentation/en-us/red_hat_enterprise_linux/6/html-single/performance_tuning_guide/index. [Accessed 11 2017].

[3] H. Servat, G. Llort, J. González, J. Giménez and J. Labarta, "Low-Overhead Detection of Memory Access Patterns and Their Time Evolution," *Euro-Par 2015: Parallel Processing: 21st International Conference on Parallel and Distributed Computing, Vienna, Austria, Proceedings,* pp. 57--69, 2015.

[4] H. Servat, G. Llort, J. Gimenez, K. Huck and J. Labarta, "Unveiling internal evolution of parallel application computation phases," *International Conference on Parallel Processing,* pp. 155-164, 2011.

[5] H. Servat, J. Labarta and J. Giménez, *Correlating Performance, Code Location and Memory Access,* Lake Tahoe, 2016.

[6] "Performance Optimisation and Productivity - A Centre of Excellence in Computing Applications," [Online]. Available: https://pop-coe.eu/. [Accessed 11 2017].

## Document History

| Version | Author(s) | Date | Changes |
|---------|-----------|------|---------|
| 0.1 | Cyril Mazauric | 2016/01 | Initial version |
| 0.2 | Erwan Raffin | 2016/02 | Perf mem tool and problem specification |
| 0.3 | Cyril Mazauric | 2016/03 | TotalView MemoryScape tool |
| 0.4 | Erwan Raffin | 2016/03 | OSS tool |
| 0.5 | Erwan Raffin | 2016/04 | ESCAPE template |
| 0.6 | Erwan Raffin | 2017/11 | BSC tools |
| 0.7 | Erwan Raffin | 2017/11 | Draft introduction, problem specification and minor modifications |
| 0.8 | Erwan Raffin | 2017/11 | Add "Application of BSC tools on an ESCAPE dwarf" as a new section. Add references. |
| 0.9 | Erwan Raffin | 2017/11 | Clarifications and improvements thanks to Oliver Fuhrer feedback |
| 0.10 | David Guibert | 2017/11 | Clarifications and improvements |
| 1.0 | Erwan Raffin | 2017/12 | Final version |

## Internal Review History

| Internal Reviewers | Date | Comments |
|--------------------|------|----------|
| Nick New (OSYS) | 15/12/2017 | Approved with comments |
| Peter Messmer (NVIDIA) | 18/12/2017 | Approved with comments |
| | | |

## Effort Contributions per Partner

| Partner | Efforts |
|---------|---------|
| Bull | 3 |
| | |
| **Total** | 3 |

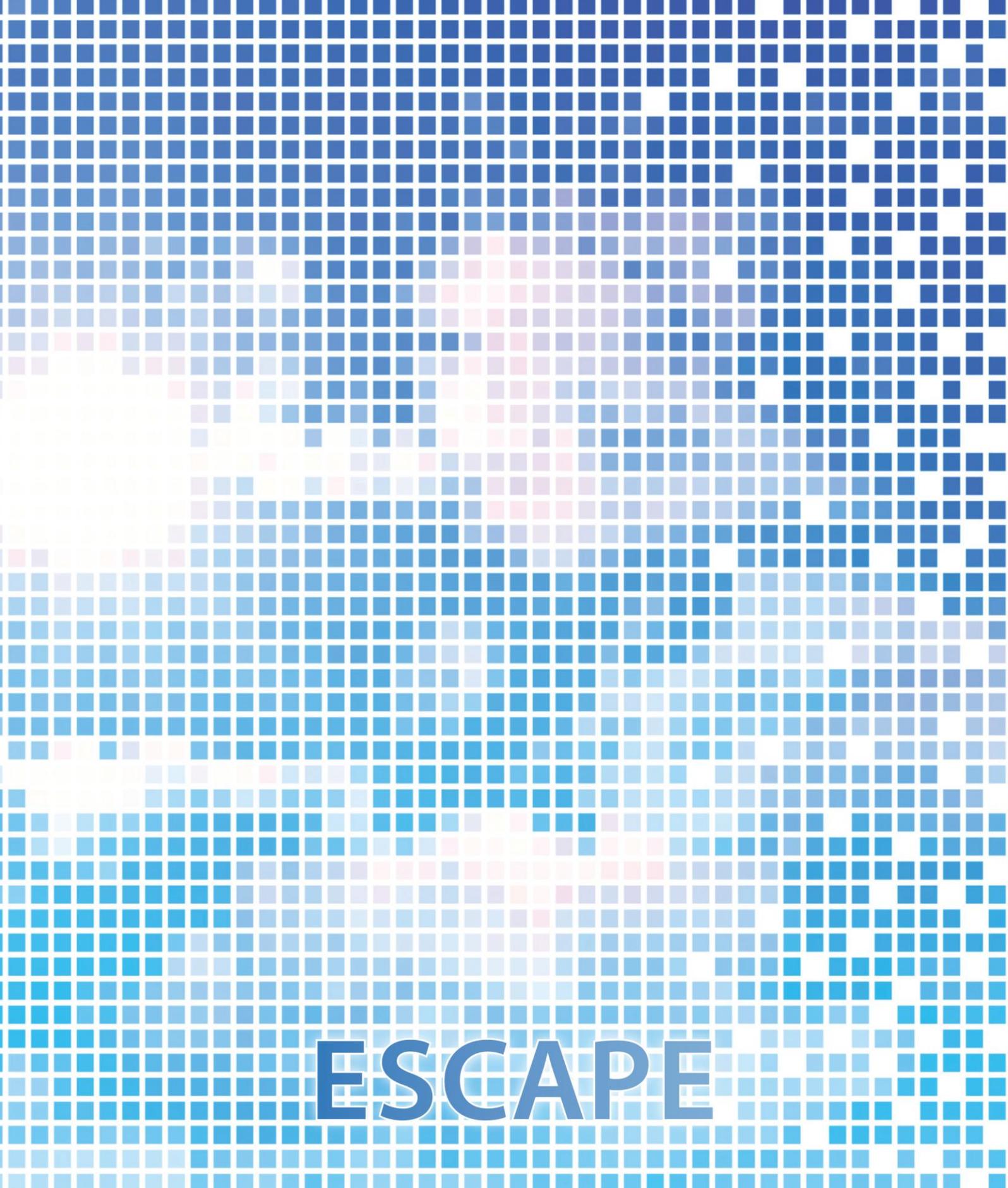